\documentstyle[11pt,epsfig]{article}
\newcommand \be{\begin{equation}}
\newcommand \ba{\begin{eqnarray}}
\newcommand \ee{\end{equation}}
\newcommand \ea{\end{eqnarray}}

\begin{document}

\begin{center}
{\LARGE The US 2000-2002 Market Descent: \\How Much Longer and Deeper?}
\end{center}
\bigskip
\begin{center}
{\large Didier
Sornette {\small$^{\mbox{\ref{igpp},\ref{ess},\ref{lpec}}}$} and Wei-Xing Zhou
{\small$^{\mbox{\ref{igpp}}}$}}
\end{center}
\bigskip
\begin{enumerate}
\item Institute of Geophysics and Planetary Physics, University of California,
Los Angeles, CA 90095\label{igpp}
\item Department of Earth and Space Sciences, University of
California, Los Angeles, CA 90095\label{ess}
\item Laboratoire de Physique de la Mati\`ere Condens\'ee, CNRS UMR 6622 and
Universit\'e de Nice-Sophia Antipolis, 06108 Nice Cedex 2, France\label{lpec}
\end{enumerate}

\date{today}

\begin{abstract}
A remarkable similarity in the behavior of the US S\&P500 index from 1996 to August 2002 and 
of the Japanese Nikkei index from 1985 to 1992 (11 years shift) is presented,
with particular emphasis on the structure of the bearish phases. Extending a
previous analysis of Johansen and Sornette [1999, 2000] on the Nikkei index ``anti-bubble''
based on a theory of cooperative herding and imitation working both in bullish
as well as in bearish regimes,
we demonstrate the existence of a clear signature of herding
in the decay of the S\&P500 index 
since August 2000 with high statistical significance, in the form of
strong log-periodic components. We offer 
a detailed analysis of what could be
the future evolution of the S\&P500 index over the next two years, according to
three versions of the theory: we expect an overall continuation of
the bearish phase, punctuated by local rallies;
we predict an overall increasing market until the end of the year 2002 or 
at the beginning of 2003 (first quarter);
we predict a strong following descent (with maybe one or two 
severe up and downs in the middle) which stops during the first 
semester of 2004. 
After this strong minimum, the market is expected
to recover. Beyond, our prediction horizon is made fuzzy by the possible
effect of additional nonlinear collective effects and of a real departure from the anti-bubble
regime.
The similarities between the two stock market
indices may reflect deeper similarities between the fundamentals of
two economies which both went through over-valuation with strong speculative phases
preceding the transition to bearish phases 
characterized by a surprising number of bad surprises (bad loans for Japan
and accounting frauds for the US) sapping investors' confidence.
\end{abstract}

\section{Introduction} \label{s1:intro}

Financial crashes have been proposed to be critical
phenomena in the statistical physics sense
of critical phase transitions
(see \cite{SJB96,CriCrash99,JSL99,CriCrash00,SorJoh01,bookcrash}
and references therein). Two hallmarks of criticality have
been documented: (i) super-exponential power law acceleration of the price
towards a ``critical'' time $t_c$ corresponding to the end of
the speculative bubble and (ii) log-periodic modulations accelerating
according to a geometric series signaling a discrete hierarchy of time scales.

Imitation between investors and their herding behavior not only
lead to speculative bubbles with accelerating overvaluations of
financial markets possibly followed by crashes, but also to
``anti-bubbles'' with decelerating market devaluations following
market peaks. There is thus a certain degree of
symmetry between the speculative behavior of the ``bull'' and
``bear'' market regimes. This degree of symmetry, after the
critical time $t_c$, corresponds to the existence of
``anti-bubbles,'' characterized by a power law decrease of the
price (or of the logarithm of the price) as a function of time
$t>t_c$, down from a maximum at $t_c$ (which is the beginning of the
anti-bubble) and by
decelerating/expanding log-periodic oscillations \cite{Nikkei99}.
The classic example of such an anti-bubble is the long-term
depression of the Japanese index, the Nikkei, that has decreased
along a downward path marked by a succession of up and downs
since its all-time high of 31 Dec. 1989. Another good
example is found for the gold future prices after 1980, after its
all-time high. The Russian market prior to and after its
speculative peak in 1997 also constitutes a remarkable example
where both bubble and anti-bubble structures appear simultaneously
for the same $t_c$. This is however a rather rare occurrence,
probably because accelerating markets with log-periodicity often
end-up in a crash, a market rupture that thus breaks down the
symmetry ($t_c-t$ for $t<t_c$ into  $t-t_c$ for $t>t_c$). Herding
behavior can occur and progressively weaken from a maximum in
``bearish'' (decreasing) market phases, even if the preceding
``bullish'' phase ending at $t_c$ was not characterized by an
imitation run-away. The symmetry is thus statistical or
global in general and holds in the ensemble rather than for each
single case individually.

In \cite{Nikkei99,JS00}, the decrease of the Nikkei index has been
analyzed in details, starting from 1 Jan. 1990, using three
increasingly complex formulas,
corresponding to the three successive orders
of a Landau expansion around $0$ of the logarithm of the price: ${d\ln F(x)
\over d\ln x} = \alpha F(x) + \cdots$, where in general the
coefficients may be complex. The first-order expression based on
discrete scale invariance \cite{SorDSI} of stock indices reads:
\be
\ln p\left( t\right) \approx A_1 + B_1 \tau ^{\alpha} +C_1
\tau^\alpha \cos\left[ \omega \ln \left( \tau \right) +\phi_1
\right], \label{Eq:fit1}
\ee
where
\be
\tau = t-t_c~,
  \label{Eq:tau1}
\ee
where $t_c$ is the time of the beginning of the anti-bubble.
The inclusion of a non-linear
quadratic term in the Landau expansion leads to the second-order
log-periodic formula \cite{SJ97}
\be
\ln p(t) \approx A_2 +
\frac{\tau^\alpha} {\sqrt{1+ \left(\frac{\tau} {\Delta_t}\right)
^{2\alpha}}} \left\{B_2+ C_2\cos\left[\omega\ln(\tau) +
\frac{\Delta_{\omega}}{2\alpha}\ln \left(1+\left(\frac{\tau}
{\Delta_t}\right)^{2\alpha}\right) +\phi_2\right]\right\}~.
\label{Eq:fit2}
\ee
A third-order formula has also been given in
\cite{Nikkei99} which derives from the addition of a third-order
term in the Landau expansion. We do not write this formula
explicitely here as we shall not need it for the present analysis.
Equation (\ref{Eq:fit2}) describes a transition from an
angular log-frequency $\omega$ (for $\tau < \Delta_t$) to 
a different angular log-frequency $\omega +
\Delta_{\omega}$ (for $\Delta_t < \tau $). Note that expression
(\ref{Eq:fit2}) reduces to equation (\ref{Eq:fit1}) in the limit
$\Delta_t \to +\infty$. Using these three formulas, a prediction
was published in January 1999 on the behavior of the Japanese
stock market in the following two years  \cite{Nikkei99}, that
have been remarkably successful \cite{JS00}.

\begin{figure}
\begin{center}
\epsfig{file=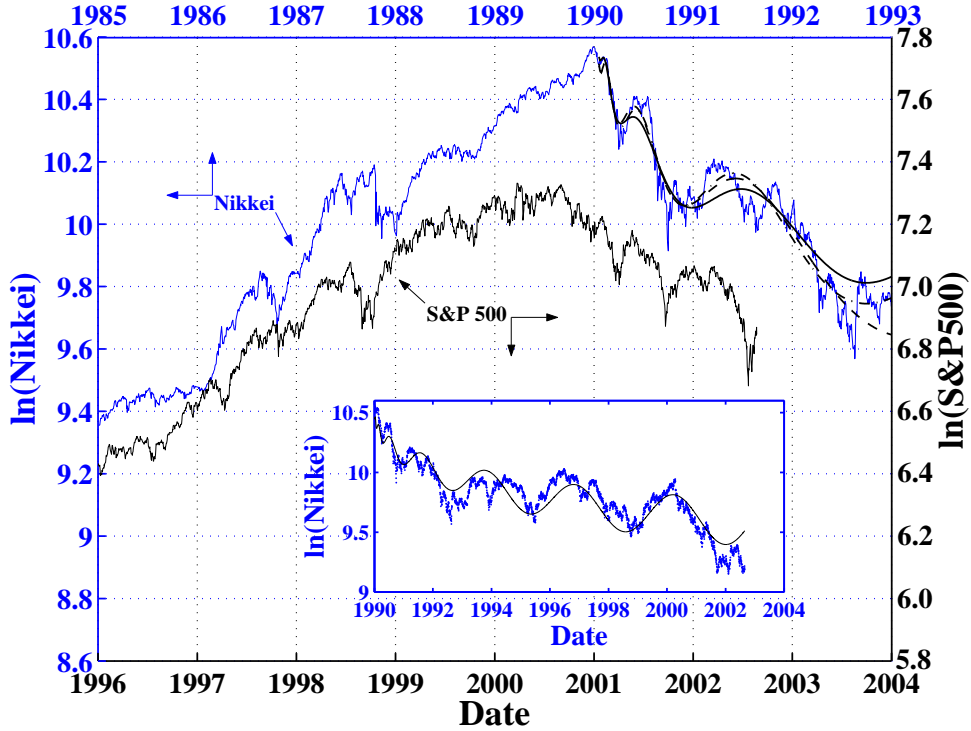,width=13cm, height=10cm}
\end{center}
\caption{Comparison between the evolutions of the US S\&P500 index
from 1996 till August, 24, 2002 (bottom and right axes) and the
Japanese Nikkei index from 1985 to 1993 (top and left axes). 
The years are written on the 
horizontal axis (and marked by a tick on the axis) where January 1
of that year occurs. The
dashed line is the simple log-periodic formula (\ref{Eq:fit1})
fitted to the Nikkei index. The data used in this fit goes from
01-Jan-1990 to 01-Jul-1992 \cite{Nikkei99}. The parameter values
are $t_c=$ 28-Dec-1989, $\alpha=0.38$, $\omega=5.0$, $\phi=2.59$,
$A=10.76$, $B=-0.067$ and $C=-0.011$. The fit error is $\chi =
0.0535$. The dash-dotted line is the improved nonlinear
log-periodic formula (\ref{Eq:fit2}) developed in \cite{SJ97}
fitted to the Nikkei index. The Nikkei index data used in this fit
goes from 01-Jan-1990 to 01-Jul-1995 \cite{Nikkei99}. The
parameter values are $t_c=$ 27-Dec-1989, $\alpha=0.38$,
$\omega=4.8$, $\phi=6.27$, $\Delta_t=6954$, $\Delta_\omega=6.5$,
$A=10.77$, $B=-0.070$, $C=0.012$. The fit error is $\chi =
0.0603$. The continuous line is the fit of the Nikkei index with
the third-order formula developed in Ref.~\cite{Nikkei99}. The
Nikkei index data used in the fit goes from 01-Jan-1990 to
31-Dec-2000. The fit is performed by fixing $t_c$, $\alpha$ and
$\omega$ at the values obtained from the second-order fit and
adjusting only $\Delta_t$, $\Delta'_t$, $\Delta_\omega$,
$\Delta'_\omega$ and $\phi$. The parameter values are
$\Delta_t=1696$, $\Delta'_t=5146$, $\Delta_\omega=-1.7$,
$\Delta'_\omega=40$, $\phi=6.27$, $A=10.86$, $B=-0.090$,
$C=-0.0095$. The fit error is $\chi = 0.0867$. In the three fits,
$A$, $B$ and $C$ are slaved to the other variables by multiplier
approach in each iteration of optimization search. The inset shows
the 13-year Nikkei anti-bubble with the fit with the third-order
formula shown as the continuous line. The parameter values are
$\Delta_t=52414$, $\Delta'_t= 17425$, $\Delta_\omega=23.7$,
$\Delta'_\omega=127.5$, $\phi=5.57$, $A=10.57$, $B=-0.045$,
$C=0.0087$. The fit error is $\chi = 0.1101$. In all our fits,
times are expressed in units of days, in contrast with the yearly
unit used in \cite{Nikkei99}). Thus, the parameters $B$ and $C$
are different since they are unit-dependent, while all the other
parameters are independent of the units. } \label{Fig:NikkeiSP}
\end{figure}

The present situation of Japan does not seem any more very
different from that of the US after the burst of the ``new
economy'' bubble in March-April 2000 \cite{nasdaqlo} (paralleling
the end of the Japanese bubble in January 1990) and the cascade of
discoveries, not yet fully unveiled in their full extent, of
creative accounting of companies striving to look good in the eyes
of analysts rather than to build strong fundamentals (paralleling
the discovery of a surprising amount of bad loans held by Japanese
banks). This remark takes a forceful meaning when looking at
Fig.~\ref{Fig:NikkeiSP}, which compares the behavior of the
Japanese Nikkei index and of the US S\&P500 index under a time
shift of 11 years. The three
fits of the Nikkei index, shown in Fig.~\ref{Fig:NikkeiSP} as
undulating lines, use the three mathematical expressions discussed
above. The dashed line is the simple log-periodic formula
(\ref{Eq:fit1}); the dotted line is the improved nonlinear
log-periodic formula (\ref{Eq:fit2}) developed in \cite{SJ97} and
also used for the October 1929 and October 1987 crashes over 8
years of data; the continuous line is the extension of the
previous nonlinear log-periodic formula to the third-order Landau
expansion developed in Ref.~\cite{Nikkei99}. This last more
sophisticated mathematical formula predicts the transition from
an angular log-frequency $\omega$ (for $\tau<\Delta_t$) to 
another angular log-frequency $\omega +
\Delta_{\omega}$ for $\Delta_t < \tau < \Delta'_{t}$ and to a third
angular log-frequency
$\omega + \Delta_{\omega} + \Delta'_{\omega}$ (for
$\Delta'_{t} < \tau$).

In the next section, we use the insight provided by the theory of 
critical herding \cite{SJB96,CriCrash99,JSL99,CriCrash00,SorJoh01,bookcrash}
to analyze the S\&P500 2000-2002 antibubble. We perform a battery of 
tests, starting with parametric fits of the index with two of the above
log-periodic power law formulas, followed by the so-called Shank's transformation
applied to characteristic times. We then present two spectral analysis,
the Lomb periodogram applied to the parametrically detrended index
and the non-parametric $(H,q)$-analysis of fractal signals. These 
different approaches complement each other and confirm the remarkably
strong presence of log-periodic structures. We also detect a 
significant second-order harmonic which provides a statistically significant
improvement of the description of the data by the theory, as tested
using the statistical theory of nested hypotheses. Section 3 offers
an analysis of what could be the future evolution of the S\&P500 index
over the next two years by comparing the predictions of three formulas.
The predictions are found to be robust and consistent. We conclude by 
speculating in 
section 4 on possible consequences as well as projections further
ahead.

\section{Analysis of the S\&P500 2000-2002 antibubble}
\label{s1:DA}

\subsection{Theoretical foundations}

The analysis presented below relies on
a general theory of financial crashes and of
stock market instabilities developed in a series of 
works (see \cite{SJB96,CriCrash99,JSL99,CriCrash00,SorJoh01,bookcrash} and
references therein). The main ingredient of the theory
is the existence of positive feedbacks in stock markets as well 
as in the economy. Positive feedbacks, i.e., 
self-reinforcement, refer to the fact that, conditioned
on the observation that the market has recently moved 
up (respectively down), this makes it more probable to keep it moving up 
(respectively down), so that a large cumulative move may ensue. 
The concept of ``positive feedbacks'' has a long
history in economics and is related to the idea of 
``increasing returns''-- which says that goods become cheaper the more of them
 are produced (and the closely related idea that some products, like fax
 machines, become more useful the more people use them). 
 ``Positive feedback'' is the opposite of ``negative feedback'', 
 a concept well-known for instance in population dynamics: the larger the population
 of rabbits in a valley, the less they have grass per rabbit. If the population
 grows too much, they will eventually starve, slowing down their reproduction rate
 which thus reduces their population at a later time.
 Thus negative feedback means that the higher the population, the 
 slower the growth rate, leading to a spontaneous regulation of 
 the population size; negative feedbacks thus tend to regulate growth
 towards an equilibrium.
 In contrast, positive feedback asserts that the 
 higher (respectively lower)
 the price or the price return in the recent past, the higher 
 (respectively lower) will be
 the price growth in the future. Positive feedbacks, when unchecked, can produce
 runaways until the deviation from equilibrium is so large that other 
 effects can be abruptly triggered and lead to rupture or crashes. Alternatively, 
 it can give prolonged depressive bearish markets.

There are many
mechanisms leading to positive feedbacks
including hedging derivatives, insurance portfolios, investors' over-confidence,
imitative behavior and herding between investors.
Such positive feedbacks
provide the fuel for the development of speculative bubbles as well as
anti-bubbles \cite{Nikkei99,JS00}, by the mechanism of cooperativity, 
that is, the interactions and imitation between traders may lead
to collective behaviors similar to crowd phenomena. 
Different types of collective regimes are separated by so-called
critical points which, in physics, are widely considered to be one of the most
interesting properties of complex systems. A system goes critical when local influences
propagate over long distances and the average state of the system becomes
exquisitely sensitive to a small perturbation, {\it i.e.} different parts of
the system become highly correlated. Another characteristic is that
critical systems are self-similar across scales: at the
critical point, an ocean of traders who are mostly bearish may have within it several
continents of traders who are mostly bullish, each of which in turns surrounds
seas of bearish traders with islands of bullish traders; the progression
continues all the way down to the smallest possible scale: a single trader \cite{Wilson}.
Intuitively speaking, critical self-similarity is why
local imitation cascades through the scales into global coordination.
Critical points are described in mathematical parlance as singularities
associated with bifurcation and catastrophe theory. At critical points,
scale invariance holds and its signature is the power law behavior of
observables. 

The last ingredient of the model is to recognize that
the stock market is made of actors which differ in
size by many orders of magnitudes ranging from individuals to gigantic
professional investors, such as pension funds. Furthermore, structures at
even higher levels, such as currency influence spheres (US\$, Euro, YEN ...),
exist and with the current globalization and de-regulation of the market
one may argue that structures on the largest possible scale, i.e.,
the world economy, are beginning to form. This means that the structure
of the financial markets have features which resembles that of hierarchical
systems with ``traders'' on all levels of the market. Of course, this
does not imply that any strict hierarchical structure of the stock market
exists, but there are numerous examples of qualitatively hierarchical
structures in society. Models of imitative interactions on
hierarchical structures predict that the power law behavior
can be decorated by so-called log-periodic corrections. Indeed,
through the existence of prefered scales in a discrete
hierarchy, or a discrete cascade of instabilities \cite{SorDSI}
or the existence of a competition between positive and negative nonlinear
feedbacks \cite{Idesor}, the scale invariance characterizing
critical points may be partially broken
into a discrete scale invariance, that is, the observable is invariant
with respect to changes of scales which are integer powers of a 
fundamental scaling ratio $\lambda$.
It is easy to show that log-periodicity
as given by the term
$C_1 \tau^\alpha \cos\left[ \omega \ln \left( \tau \right) +\phi_1
\right]$ of expression (\ref{Eq:fit1}) is the signature of discrete
scale invariance: the term $ \cos\left[ \omega \ln \left( \tau \right) +\phi_1
\right]$ reproduces itself each time $\ln \tau$ changed by $2 \pi/\omega$, 
that is, each time $\tau$ is multiplied by $\lambda=\exp[2 \pi/\omega]$.
This theory predicts 
robust and universal signatures of speculative phases of financial markets,
both in accelerating bubbles as well as in decelerating anti-bubbles.
These precursory patterns have
been documented for essentially all crashes on developed as well as emergent
stock markets, on currency markets, on company stocks, etc.

\subsection{Log-periodic fits \label{lpana}}

We use equations (\ref{Eq:fit1}) and (\ref{Eq:fit2}) to fit the
logarithm of the S\&P500 index over
an interval starting at time $t_{\rm
start}$ and ending at Aug. 24, 2002. The justification of the use of the
logarithm of the price is presented in \cite{CriCrash99}. The
choice of $t_{\rm start}$ is not completely obvious. It is clear
that it should be close to the peak of the S\&P500 index in 2000
but cannot be expected to be exactly coincident with the time of
the peak due to finite-size and other effects spoiling the
validity of the log-periodic power law. We address this problem in
two ways. First, we scan $t_{\rm start}$ and select 10 time
series, starting respectively at $t_{\rm start}=1$st March 2000,
$1$st April 2000, ..., $1$st December 2000. The comparison of the
fits obtained for these 10 time series will give a sense of their
sensitivity with respect to $t_{\rm start}$. Second, we notice
that we can generalize the definition of $\tau$ given by
(\ref{Eq:tau1}) into
\be
\tau=|t-t_c|~.
\label{Eq:tau3}
\ee
While
definition (\ref{Eq:tau1}) together with the logarithmic as well
as power law singularities associated with formulas
(\ref{Eq:fit1}) and (\ref{Eq:fit2}) imposes that $t_c < t_{\rm
start}$ for an anti-bubble, the definition (\ref{Eq:tau3}) allows
for the critical time $t_c$ to lie anywhere within the time
series. In that case, the part of the time series for $t<t_c$
corresponds to an accelerating ``bubble'' phase while the part
$t>t_c$ corresponds to a decelerating ``anti-bubble'' phase.
Definition (\ref{Eq:tau3}) has thus the advantage of introducing a
degree of flexibility in the search space for $t_c$ without much
additional cost. In particular, it allows us to avoid a thorough
scanning of $t_{\rm start}$ since the value of $t_c$ obtained with
this procedure is automatically adjusted without constraint.

There are two potential problems associate with this new procedure
(\ref{Eq:tau3}). First, it assumes that the anti-bubble is always
associated with a bubble which, in addition, has the same $t_c$.
Second, it assumes that the bubble and anti-bubble are exactly
symmetric around $t_c$, that is, the same parameters characterize
the index evolution for $t<t_c$ and for $t>t_c$. For the cases
relevant to the present study, these two problems are quite minor
and can be neglected because $t_c$ is always found close to
$t_{\rm start}$ (within the time series for $t_{\rm start}$ prior
to August 2000 and anterior to the time series otherwise). Our
comparison with fits using (\ref{Eq:tau1}) shows that the new
procedure provides significantly better and more stable
fits, with in particular a value of $t_c$ very weakly sensitive to
$t_{\rm start}$. The parameters of the fits
with the first-order formula using (\ref{Eq:tau1}) or
(\ref{Eq:tau3}) are presented in Table
\ref{Tb}, using subscripts 0 and 1 respectively. The
fits with the first-order formula (\ref{Eq:fit1})
with the definition (\ref{Eq:tau1}) are unstable, are quite
sensitive to $t_{\rm{start}}$ and have on average much larger
values of their residuals $\chi$ ($\chi_0>\chi_1$). Comparing parameters with
subscripts 1 and 2, one can see that applying the
symmetry condition (\ref{Eq:tau3})
improves the quality of fit remarkably. We stress however that
this improvement (\ref{Eq:tau3}) is not crucial and our results reported
below remain robust with the definition (\ref{Eq:tau1}). It
would also be quite easy to relax the constraints of
(\ref{Eq:tau3}) by replacing $\tau=|t-t_c|$ by an asymmetric
function allowing in addition for a plateau or time lag. Since
this would involve additional and poorly constrained additional
parameters, we do not pursue this possibility here.

The results of the fits of the logarithm of the S\&P 500
index from $t_{\mathtt{start}}$ to August, 24, 2002 with Eqs.
(\ref{Eq:fit1}) and (\ref{Eq:fit2}) using the improved scheme
(\ref{Eq:tau3}) are presented in figure \ref{Fig:all}
and in Table \ref{Tb} under the subscripts
1 and 2 respectively. The ten oscillating curves correspond to the ten
best fits, one for each of the 10 chosen values of $t_{\rm start}$ from
March to December 2000.
Over the approximately two years period
available for the S\&P500 anti-bubble, we find that the two
formulas give essentially the same results and the same
predictions for the following year. This is reflected
quantitatively by the facts that $\chi_1 \approx \chi_2$ and that
the parameters $\Delta_t$'s are extremely large, in which case expression
(\ref{Eq:fit2}) reduces to equation (\ref{Eq:fit1}) in the limit
$\Delta_t \to +\infty$. The top (respectively bottom) panel
corresponds to using equation (\ref{Eq:fit1}) (respectively
(\ref{Eq:fit2})).  The curves are shown as continuous line in
their fitting interval and as dotted line in their extrapolation
to the future. Note the very robust nature of the solutions
obtained for the ten choices of  $t_{\mathtt{start}}$, which
essentially all agree in their parameters and in
their prediction of the future evolution.

To sum up, varying the starting date $t_{\rm start}$ of the fitted
time window over a 10-month period and using two formulas, we
confirm that a single log-periodic power law describes very well
the S\&P500 anti-bubble since around mid-2000
According to the values of $t_{c,1}$ listed in Table \ref{Tb}, the
critical $t_c$ is around 09-Aug-2000. This is consistent with the
fact that the fit residuals obtained with $t_{\rm{start}} = $ 01-Mar-2000,
01-Apr-2000, 01-Oct-2000, 01-Nov-2000 and 01-Dec-2000 are
significantly larger ($\chi_1> 3.3$) than the residuals for the other
values of $t_{\rm{start}}$. It is natural that
fits with $t_{\rm{start}}$ close to $t_c$ have smaller fit
residuals. The fits with $t_{\rm{start}} = $ 01-May-2000,
01-Jun-2000, 01-Jul-2000, 01-Aug-2000 and 01-Sep-2000 indeed exhibut
smaller residuals. This conclusion is also supported by the values of
$\omega$ and
of $\alpha$ which are basically constant for these five
starting dates. Combining all the information shown in Table \ref{Tb},
we have the critical time $t_c = $ 09-Aug-2000
$\pm 5$ days, the angular log-frequency $\omega = 10.30 \pm 0.17$
and the exponent $\alpha = 0.69 \pm 0.02$.
The fits with the second-order formula (\ref{Eq:fit2}) gives similar results.

\subsection{Analysis using the Shank's transformation on a hierarchy
of characteristic times}

The fundamental idea behind the appearance of log-periodicity is
the existence of a hierarchy of characteristic scales.
Reciprocally, any log-periodic pattern implies the existence of a
hierarchy of characteristic time scales. This hierarchy of time
scales is determined by the local positive maxima or minima of the
function such as $\log[p(t)]$. For the S\&P500 index anti-bubble,
let us consider the times $t_n$ at which the S\&P500 index reached
a local minimum. As seen in figure \ref{Fig:all}, there is a clear
sequence of sharp minima. We number them from the most recent one
$t_1=$ 23-Jul-2002, $t_2=$ 21-Sep-2001, $t_3=$ 04-Apr-2001, $t_4=$
20-Dec-2000 up to the earliest one at $t_5=$ 12-Oct-2000, which is
still obvious. According to the prediction of log-periodicity, the
spacing between successive values of $t_n$ approaches zero as a
geometric series as $n$ becomes large and $t_n$ converges to
$t_c$. We have $t_1-t_2=305$ days, $t_2-t_3=170$ days,
$t_3-t_4=105$ days and $t_4-t_5=69$ days.

Specifically, log-periodicity predicts that the times $t_n$ are
organized in a geometric time series such that
\be
t_n - t_c =
{\tau \over \lambda^n}~,
\label{jkslls}
\ee
where $\tau$ sets the
time unit and
\be
\lambda = e^{2\pi \over \omega}~
\ee
is the prefered scaling ratio. The
relation (\ref{jkslls}) leads to
\be
{t_n - t_c \over t_{n+1} -
t_c} = {t_n - t_{n+1} \over t_{n+1} - t_{n+2}} = \lambda~,
\label{ratio}
\ee
which is a signature of the discrete self-similarity of the log-periodic
oscillations.
Using the previously determined dates $t_1, ...,
t_5$, we obtain
\be
{t_1 - t_2 \over t_2 - t_3}= 1.79 ~,
\label{1}
\ee
\be
{t_2 - t_3 \over t_3 - t_4}= 1.62~,
\label{2}
\ee
\be
{t_3 - t_4 \over t_4 - t_5}= 1.52~.
\label{3}
\ee
These three values
given by (\ref{1},\ref{2},\ref{3}) are compatible but smaller
than the value of $\lambda =1.9 \pm 0.1$ deduced from the log-periodic fits
shown in Table \ref{Tb}. Note that, according to the theory,
the hierarchy of scales $t_c - t_n$ are not universal but depend
upon the specification of the system. What is expected to be
universal are the ratios $\frac{t_{n+1}-t_c }{t_n-t_c} = \lambda$.

 From three successive observed values of $t_n$, say $t_n$,
$t_{n+1}$ and $t_{n+2}$, we can obtain an estimation of the
critical time by the following formula \be t_c = {t_{n+1}^2 -
t_{n+2}t_n \over 2t_{n+1} - t_n - t_{n+2}}~ . \label{ert} \ee This
relation applies the so-called Shanks transformation \cite{Bender}
to accelerate the convergence of series. In the case of an exact
geometrical series, three terms are enough to converge exactly to
the asymptotic value $t_c$. Notice that this relation is invariant
with respect to an arbitrary translation in time. Applying
(\ref{ert}) with $t_1, t_2, t_3$ gives $t_c=$ 02-Sep-2000.
Applying (\ref{ert}) with $t_2, t_3, t_4$ gives $t_c=$
03-Jul-2000. Applying (\ref{ert}) with $t_3, t_4, t_5$ gives
$t_c=$ 01-Jun-2000. These back-predictions for $t_c$ are compatible
with the value of $t_c = $ 09-Aug-2000 $\pm 5$ days
given in Table \ref{Tb} determined by the log-periodic fits.

In addition, the geometric structure of the time series $t_1, t_2,
\cdots$ is such that the next time $t_{n+3}$ can be obtained from
the first three ones by \be t_{n+3} = {t_{n+1}^2 + t_{n+2}^2 - t_n
t_{n+2} - t_{n+1} t_{n+2} \over t_{n+1} - t_n}~ . \label{sdfbis}
\ee Since time is measured backward as $n$ increases, we are
interested in the time $t_0$ at which the next future minimum
occurs. For this, we use formula (\ref{sdfbis}) and put $n=0$ to
get \be t_0 = { t_1^2+t_2^2 -t_1t_2 - t_1t_3 \over t_2-t_3}
\approx {\rm{21~Jan~2004}}. \label{Eq:ShankPred}\ee

While this Shank analysis has the advantage of simplicity and
of having an obvious geometrical interpretation, its
weakness lies in using a geometric set
of characteristic times $\{t_n\}$ whose identification
may be quite subjective. In the
present case, the minima are so distinctive and sharp that there
is no ambiguity.
But,the use of only three characteristics dates makes more
serious the sensitivity to noise and
is of course bounded
to lead to less precise fits and predictions that the full
parametric fits reported in the previous section \ref{lpana}.

\subsection{Spectral analysis}

While the two previous analyses are suggestive, the parametric
nature of the first one and the limited power of the second one
requires additional tests of the reported log-periodicity. In this
goal, we now turn to objective approaches for the detection of
log-periodicity by applying a spectral Lomb analysis \cite{Press}.
The Lomb analysis is a spectral analysis designed for unevenly
sampled data which gives the same results as the standard Fourier
spectral analysis. We apply this spectral analysis to two types of
signals. Following \cite{JohSorLed99}, the first one is obtained
by detrending the logarithm of the S\&P500 index, using the power
law (without log-periodicity) with the exponent $\alpha$
determined from the previous fits with formula (\ref{Eq:fit1}).
The time series of the residuals of the simple power law fit
should be a pure cosine of $\ln \tau$ if log-periodicity was
perfect. We also use a recently developed non-parametric approach
called the $(H,q)$-analysis, which has been successfully applied
to financial crashes \cite{HqACrash} and critical ruptures
\cite{HqARup} for the detection of log-periodic components.

\subsubsection{Parametric detrending approach \label{paradet}}

We first construct the following detrended quantity which is
defined using Eq.~(\ref{Eq:fit1}) by:
\be
r(t) = \ln
\left[{A_1-\ln p(t) \over (t-t_c)^\alpha}\right] - \left\langle\ln
\left[{A_1-\ln p(t) \over (t-t_c)^\alpha}\right]\right\rangle~,
\label{Eq:r}
\ee
where the bracket refers to the sample average.
Table \ref{Tb} shows that the values of
$A_1$ and $\alpha$ are quite stable and approximately equal to $7.33$
and $0.69$
for different $t_{\rm{start}}$. In order to investigate the impact of the
different choice of $t_c$, we first construct $\ln p(t)
- A_1$ as a function of $\ln(t-tc)$ for different $t_c$ by fixing
$A_1 = 7.33$. We then detrend it directly by determining the exponent $\alpha$
from a linear fit, whose residuals $r(t)$ define the time series to be
analyzed for log-periodicity.

Figure \ref{Fig:2311} shows the residual time series $r(t)$ as a
function of $\ln(t-t_c)$ for three different critical times $t_c$:
15-Jul-2000 (top panel), 01-Aug-2000 (mid panel) and 15-Aug-2000
(bottom panel). The log-periodic undulations are clearly visible.
We then perform the spectral Lomb analysis on these residuals. The
three corresponding Lomb periodograms shown in Fig.~\ref{Fig:2312}
are very consistent with each other. They all exhibit an extremely
strong and significant peak close to the log-frequency
$f=\omega/2\pi=1.7$ with an amplitude larger than 140. Its
harmonic shown as the downward pointing arrow is significant. As
we shall discuss later in section \ref{roleharm}, the existence of
harmonics have been shown to be important factors for qualifying
log-periodicity \cite{turb1,turb2}. The presence of this harmonic
is thus taken as a confirmation of the presence of
log-periodicity. The third peak at $f=2.5-3$ is also
significant but less well-constrainted. The inset of
Fig.~\ref{Fig:2312} magnifies the Lomb periodogram in the
neighborhood of the largest peaks. The two highest Lomb peaks for
$t_c=$ 01-Aug-2000 and $t_c=$ 15-Aug-2000 are significantly higher
than the highest peak for $t_c=$ 15-Jul-2000, confirming that the
critical time is somewhere between 01-Aug-2000 and 15-Aug-2000
(recall that section \ref{lpana} found $t_c \approx $ 09-Aug-2000
$\pm 5$ days). As the number of the data points used for the Lomb
analysis is the same in all analyses throughout this paper, this
warrants a comparison of the Lomb peak heights for different
$t_c$'s to establish their significance levels \cite{Zhou02a}. The
log-periodic frequency of the largest peak for $t_c=$ 01-Aug-2000
is $f = 1.71$, corresponding to $\omega = 2\pi f = 10.7$, which is
in reasonable agreement with the value $\omega = 10.30\pm 0.17$
reported in Table \ref{Tb}.

To further assess the sensitivity with respect to the choice of
$t_c$, we choose 21 different $t_c$ evenly spaced in the time
interval from 15-Jul-2000 to 13-Sep-2000. For each $t_c$, we
perform the same analysis as above and obtain the highest Lomb
peak and its associated log-frequency. The results are shown in
Fig.~\ref{Fig:2313}. The log-frequency is found to decrease with
$t_c$. This is due to the fact that, moving the critical time
forward (respectively backward), the other end of the time
series will decelerate (respectively accelerate) the log-periodic
oscillations. Fig.~\ref{Fig:2313} confirms that the best critical
time $t_c$ falls somewhere in the first half of August 2000. The
corresponding log-frequencies are $f = 1.63 - 1.72$ ($\omega =
10.2 - 10.7$), in agreement with Table \ref{Tb}.

It is interesting that the most
relevant angular log-frequency $\omega \approx 10$ fitted on the S\&P500
index is found very close to twice the
value $\approx 5$ found previously for the Nikkei index. Actually,
a small but noticeable peak at this value $f \approx 5/2\pi \approx 0.80$ 
can be seen in Fig.~\ref{Fig:2312}, strengthening the analogy quantitatively.
It may thus be surmised that both the Nikkei and the S\&P500 indices are
characterized by a universal discrete hierarchy of (angular) log-periodic
frequencies, all harmonics of a fundamental angular log-frequency close to $5$.
Other systems have previously exhibited the curious fact, also observed
when comparing the S\&P500 to the Nikkei indices, that the higher-order 
harmonics may have an amplitude larger than the fundamental value 
\cite{JSH,turb1,turb2}, as observed for the S\&P500 index.

\subsubsection{Non-parametric $(H,q)$-analysis}

The $(H,q)$-analysis \cite{HqARup,HqACrash} is a generalization of
the $q$-analysis \cite{Erzan,ErzEck}, which is a natural tool for
the description of discretely scale invariant fractals. The
$(H,q)$-derivative is defined as
\be
D_q^H f(x)
\stackrel{\triangle}{=} \frac {f(x)-f(qx)}{[(1-q)x]^H}~.
\label{Eq:HqD}
\ee
The special case $H=1$ recovers the normal
$q$-derivative, which itself reduces to the normal derivative in the limit
$q \to 1^-$.
There is no loss of generality by constraining $q$ in the open
interval $(0,1)$ \cite{HqARup}. We apply the $(H,q)$-analysis to
verify non-parametrically the existence of log-periodicity by
taking $f(x) = \ln p(t)$ and $x = t-t_c$. The advantage of the
$(H,q)$-analysis  is that there is no need of detrending as done in the
previous section \ref{paradet}. Such detrending is automatically accounted
for by the finite difference and the normalization by the denominor.

Fig.~\ref{Fig:2321} shows the $(H,q)$-derivatives of the logarithm
of S\&P500 index as a function f $\ln(t-t_c)$ for three choice of
$t_c$: 15-Jul-2000 with $H = 0$ and $q = 0.8$ in the top panel,
02-Aug-2000 with $H = 0.5$ and $q = 0.7$ in the mid panel, and
16-Aug-2000 with $H = 0.4$ and $q = 0.7$ in the bottom panel. Each
of these pairs of $(H,q)$ is the optimal pair \cite{HqARup,
HqACrash} corresponding to the most significant peak among all
Lomb periodograms for each $t_c$. The log-periodic undulations are
clearly visible to the naked eyes. The Lomb periodograms of the
$(H,q)$-derivatives shown in Fig.~\ref{Fig:2321} are presented in
Fig.~\ref{Fig:2322}. The three highest Lomb peaks are even more
significant than those in Fig.~\ref{Fig:2312}. The inset shows the
magnified Lomb peaks. The highest Lomb peak is obtained for $t_c=$
02-Aug-2000, which corresponds to the log-frequency $f \approx
1.71$.

To test the robustness of the $(H,q)$-analysis for the detection
of log-periodicity, we scan $H$ from $-1$ to 1 with spacing $0.1$
and $q$ from $0.1$ to $0.9$ with spacing $0.1$ for each critical
$t_c$. Figure \ref{Fig:2323} shows the log-frequency $f$ as a
function of $H$ and $q$ for $t_c=$ 16-Aug-2000. The existence of a
flat plateau at $f = 1.62 \pm 0.07$ for most of the pairs $(H,q)$
confirm the existence of log-periodicity. The five smaller
log-frequencies below the plateau correspond to the spurious
values stemming from the most probable effect of noise on power
laws \cite{fmp} and should be discarded. The three higher
log-frequencies above the plateau probably stem from the
interaction between high-frequency noise and the second harmonics.

We then apply the non-parametric $(H,q)$-analysis to the
S\&P500 index for 21 different choices of the critical $t_c$. For
each given $t_c$, we take the average of the Lomb
periodograms for all $21\times 9$ pairs of $(H,q)$. The amplitude
of the highest peak in each averaged Lomb periodogram is plotted
as a function of $t_c$ in the lower panel of Fig.~\ref{Fig:2324}.
Their associated log-frequency shown in the upper panel of
Fig.~\ref{Fig:2324} with error bars is estimated from
the average height of the plateaus such as the one
seen in Fig.~\ref{Fig:2323}. This log-frequency
slightly decreases with $t_c$. There are two clear humps in the
lower panel around late July and mid August 2000. The hump around
mid August is higher than the other one. The corresponding
log-frequencies $f \approx 1.60 - 1.70$ are compatible with those
reported in Table \ref{Tb}.

\subsection{Role of log-periodic harmonics \label{roleharm}}

The spectral Lomb analyses reported above (see figures
\ref{Fig:2312} and \ref{Fig:2322}) as well as a the visual
structure of the S\&P500 time series suggest the presence of a
rather strong harmonic at the angular log-frequency $2\omega$. The
possible importance of harmonics in order to qualify
log-periodicity is made also more credible by recent analyses of
log-periodicity in hydrodynamic turbulence data \cite{turb1,turb2}
which have demonstrated the important role of higher harmonics in
the detection of log-periodicity.

We thus revisit the parametric log-periodic fits of section
\ref{lpana} with formula (\ref{Eq:fit1}) using (\ref{Eq:tau3}) to
include the effect of an harmonic at the angular log-frequency $2
\omega$. In this goal, we postulate the formula \be \ln p\left(
t\right) \approx A + B \tau ^{\alpha} +C \tau^\alpha \cos\left[
\omega \ln \left( \tau \right) +\phi_1 \right] + D \tau^\alpha
\cos\left[ 2\omega \ln \left( \tau \right) +\phi_2 \right]~,
\label{Eq:fit4} \ee which differs from equation (\ref{Eq:fit1}) by
the addition of the last term  proportional to the amplitude $D$.
This formula has two additional parameters compared with
(\ref{Eq:fit1}), the amplitude $D$ of the harmonic and its phase
$\phi_2$. We follow the fit procedure of Ref.~\cite{BubbleTests}
with minor modifications, which itself adapts the slaving method
of \cite{JohSorLed99,CriCrash00}. By rewriting Eq.~(\ref{Eq:fit4})
as $\ln p(t) = A + B f(t) + C g(t) + D h(t)$, we obtain a system
of 4 linear equations for the four variables $A$, $B$, $C$ and
$D$:
\begin{equation}
\left( \begin{array}{l} \sum \ln p_i \\ \sum (\ln p_i) f_i
\\ \sum (\ln p_i) g_i \\ \sum (\ln p_i) h_i
\end{array} \right) = \left(
\begin{array}{llll} N & \sum f_i & \sum g_i & \sum h_i \\ \sum f_i &
\sum f_i^2 & \sum g_if_i & \sum h_if_i \\\sum g_i & \sum f_ig_i &
\sum g_i^2 & \sum h_ig_i
\\\sum h_i & \sum f_ih_i & \sum g_ih_i & \sum
h_i^2 \end{array} \right) \cdot \left(
\begin{array}{c} A\\B\\C\\D  \end{array}\right)~,
\label{Eq:ABCD}
\end{equation}
where $p_i = p(t_i)$, $f_i = f(t_i)$, $g_i = g(t_i)$ and $h_i =
h(t_i)$. Solving analytically this system allows us to slave the
four parameters $A$, $B$, $C$ and $D$ to the other parameters in
the search for the best fit. With this approach, we find that the
search of the optimal parameters is very stable and provides fits
of very good quality in spite of the remaining five free
parameters.

The results are listed in Table \ref{Tb2} and depicted in figure
\ref{Fig:AllFit4}. The fit residuals are reduced considerably
compared with the fits reported in Table \ref{Tb}. The improvement
of the fits are obvious when comparing Fig.~\ref{Fig:AllFit4} with
figure \ref{Fig:all}. Note also that $|D| < |C|$ ($|C|
\approx 5 |D|$) which is in agreement with the fact that the spectral
peak of the fundamental log-frequency is much higher than the peak
of the second harmonic approximately by a factor $5$, as shown in
Fig.~\ref{Fig:2312} and
Fig.~\ref{Fig:2322}.

We have also tested whether the addition of a third log-frequency around
$f \approx 2.8$ (which is not a third harmonic), 
as suggested from the spectral Lomb analyses shown
in figures \ref{Fig:2312} and \ref{Fig:2322}), could improve and/or modify
the fit. We found a slight but non-significant
reduction of the root-mean-square error with negligible modification of the fit,
suggesting that this log-frequency is due to noise.

Since expression (\ref{Eq:fit4}) contains the formula
(\ref{Eq:fit1}) as a special case $D=0$, we can use Wilk's theorem
\cite{Rao} and the statistical methodology of nested hypotheses to
assess whether the hypothesis that $D=0$ can be rejected.
Therefore, the null hypothesis and its alternative are
\begin{enumerate}
\item{$H_0$: $D=0$;}
\item{$H_1$: $D\ne0$;}
\end{enumerate}
The method proceeds as follows. By assuming a Gaussian
distribution of observation errors (residuals) at each data point, the maximum
likelihood estimation of the parameters amounts exactly to the
minimization of the sum of the square over all 
data points (of number $n$) of the differences $\delta_j(i)$ between the mathematical
formula and the data \cite{Press}. The standard deviation
$\sigma_j$ with $j=0,1$ of the fits to the data associated respectively
with (\ref{Eq:fit1}) and (\ref{Eq:fit4}) is given by $1/n$ times the sum
of the squares over all 
data points of the differences $\delta_j^{(o)}(i)$ between the mathematical
formula and the data, estimated for the optimal parameters of the fit.
The log-likelihoods corresponding
to the two hypotheses are thus given by
\be 
L_j = -n\ln\sqrt{2\pi}-n\ln\sigma_j - n/2~,
\ee 
where the third term results from the product of 
Gaussians in the likelihood, which is of the form 
$$
\propto \prod_{i=1}^n 
\exp[-(\delta_j^{(o)}(i))^2/2\sigma_j^2] = \exp[-n/2]~,
$$
 from the 
definition $\sigma_j^2 = (1/n) \sum_{i=1}^n [\delta_j^{(o)}(i)]^2$.
Then, according to Wilk theorem of nested hypotheses, the
log-likelihood-ratio 
\be 
T = -2 (L_0-L_1) = 2n(\ln\sigma_1 -
\ln\sigma_0)~, 
\label{Eq:T} 
\ee 
is a chi-square variable with $k$
degrees of freedom, where $k$ is the number of restricted
parameters \cite{Holden}. In the present case, we have $k=1$.

The Wilk test thus amounts to calculate the probability that the
obtained value of $T$ can be overpassed by chance alone. If this
probability is small, this means that chance is not a convincing
explanation for the large value of $T$ which becomes meaningful.
This implies a rejection of the hypothesis that $D=0$ is
sufficient to explain the data and favor the fit with $D \neq 0$
as statistically significant.

In our test, the beginning of the fitted data set is fixed at
09-Aug-2000, while the end of the data set varies from 01-Jan-2001
to 24-Aug-2002. The results of the Wilk test are presented in
Table \ref{Tb3}. Increasing the number of points decreases the
probability that the obtained probability to overpass $T$ may
result from chance, and thus increases the statistical
significance of the fit with Eq.~(\ref{Eq:fit4}). Since the
assumption of Gaussian noise is most probably an under-estimation of the
real distribution of noise amplitudes, the very significant improvement in
the quality of the fit brought by the use of Eq.~(\ref{Eq:fit4})
quantified in Table \ref{Tb3} provides most probably a lower bound
for the statistical significance of the hypothesis that $D$
should be chosen non-zero, above the $99.8\%$ confidence level.
Indeed, a non-Gaussian noise with a fat-tailed distribution would
be expected to decrease the relevance of competing formulas, whose
performance could  be scrambled and be made fuzzy. The clear and
strong result of the Wilk test with assumed Gaussion noises
thus confirm a very strong significance
of Eq.~(\ref{Eq:fit4}).

Strengthened by this analysis of the strong
relevance of the second
harmonics at $2\omega$, we revisit the Nikkei index and fit it
in the first 2.6 years of its decay starting in January 1990
using (\ref{Eq:fit4}) to test
whether the second harmonics is also important for the Nikkei index. The fit is
compared with the log-price in Fig.~\ref{Fig:Nikkei2}. We find indeed
an impressive improvement, as the mean square
error $\chi=0.0457$ is significantly smaller than the mean square error
$\chi=0.0535$ obtained with expression
(\ref{Eq:fit1}) and whose fit is shown in Fig.~\ref{Fig:NikkeiSP}.

\section{Discussion and prediction \label{s1:discu}}

Starting from a visual analogy with the Nikkei index shifted by 11
years, the first point of the analyses presented above is to have
established with strong significance the existence of an
anti-bubble followed by the S\&P500 index approximately since
July-August 2000. This anti-bubble is characterized by an overall
power law decay of the index decorated by strong log-periodic
oscillations. 

Following the analogy with the trajectory of the Nikkei index 11
years earlier, the second point is that a comparison between the
fits obtained with equations (\ref{Eq:fit1}) and (\ref{Eq:fit2})
shows that the S\&P500 index has not yet entered into the second
phase in which the angular log-frequency may start its shift to
another value, as did the Nikkei index after about 2.5 years of
its decay. We may expect this to occur in the future. Not being
able to estimate directly the parameter $\Delta_t$ controlling
this transition due to the smallness of the S\&P500 index
anti-bubble duration, we can however offer the following guess,
based on the hypothesis that the values $\Delta_t$ and
$\Delta_{\omega}$ given by the fit of expression (\ref{Eq:fit2})
to the logarithm of the Nikkei index are reasonable estimates of
those for the S\&P500 index. We use also the parameters in the
column of $t_{\rm{start}}=$ Aug-01-2000 of Table \ref{Tb} for the
first order regime and extrapolate the fitted curve to 2006
(continuous line). Using the values of the fits with
(\ref{Eq:fit1}) and plugging in the values of $\Delta_t$ and
$\Delta_{\omega}$ from the Nikkei index in expression
(\ref{Eq:fit2}) gives the dashed line shown in Fig.~\ref{Fig:301}.
The crossover from the first order regime to the second order
regime is here suggested to occur in the first half of 2004.
Fig.~\ref{Fig:301} also compares the first-order fit and the
second-order guess with formula (\ref{Eq:fit4}) taking into
account the second harmonics. Fig.~\ref{Fig:301} suggests that the
next broad minimum of the S\&P500 index will occur in the first
semester of 2004. This is consistent with the prediction
(\ref{Eq:ShankPred}) using Shank's transformation.

\begin{figure}
\begin{center}
\epsfig{file=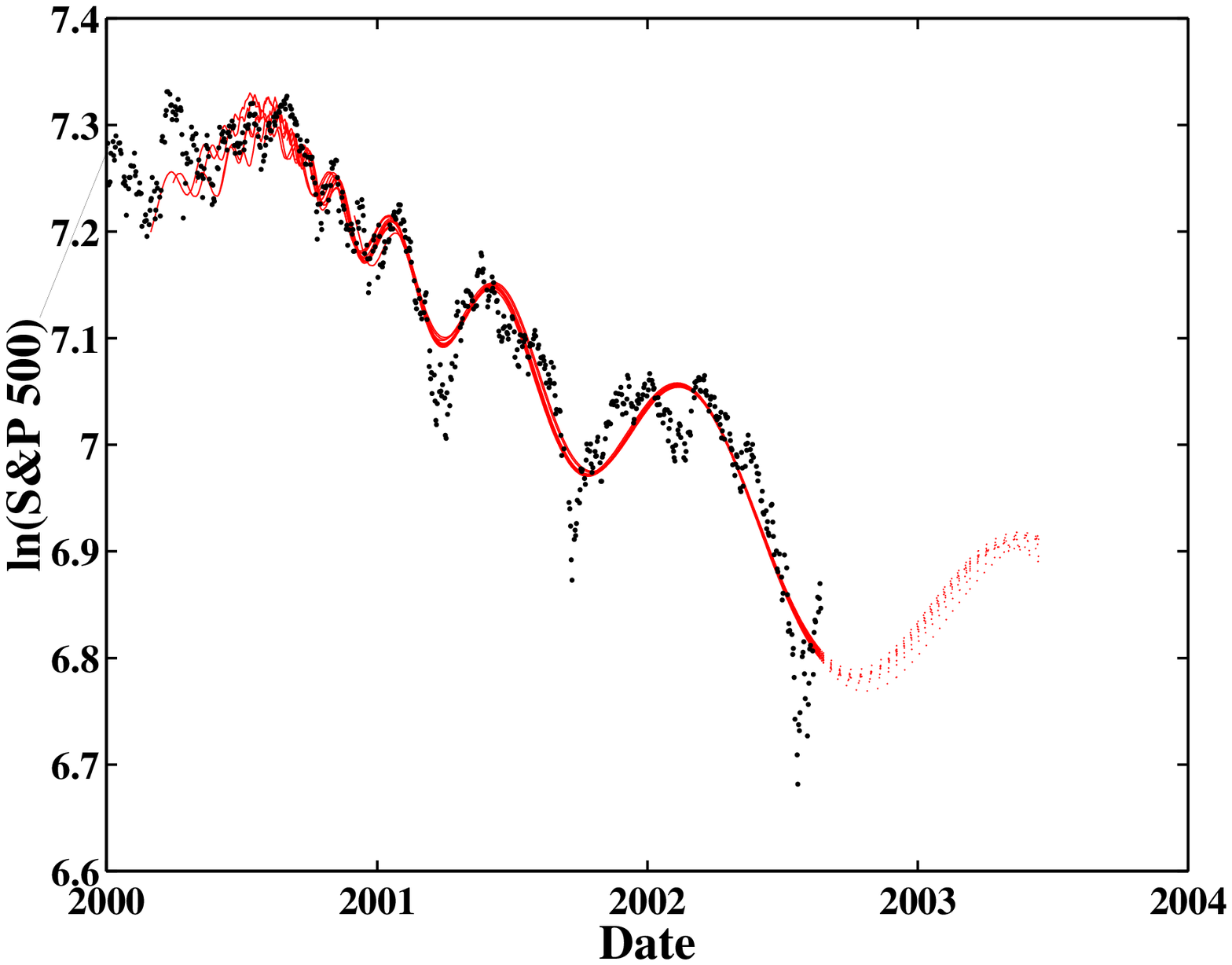,width=10cm, height=8cm}
\epsfig{file=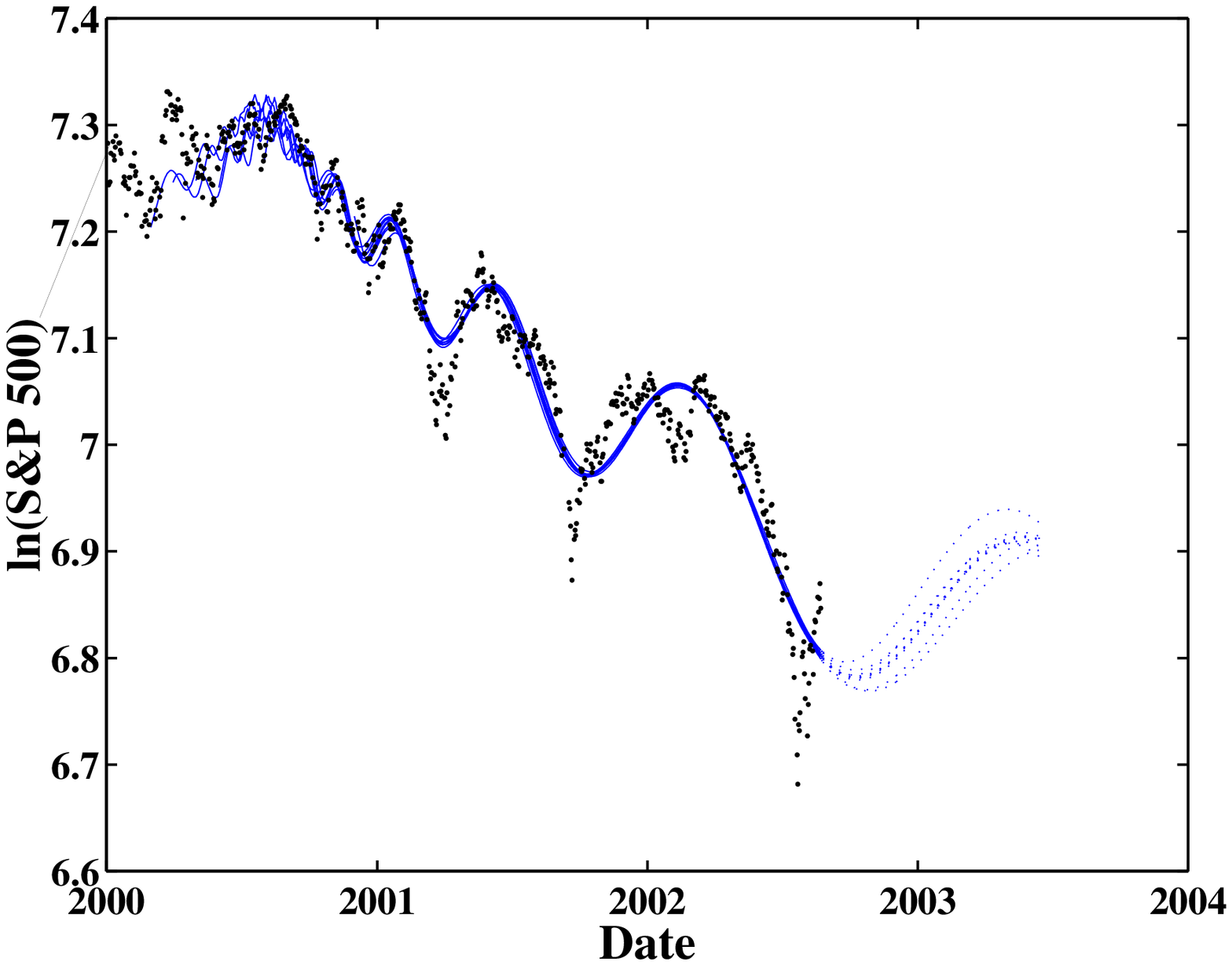,width=10cm, height=8cm}
\end{center}
\caption{The S\&P500 index anti-bubble fitted
from $t_{\rm start}$ to August, 24, 2002,
with  the improved scheme (\ref{Eq:tau3}) inserted in the two
formulas (\ref{Eq:fit1}) (upper panel) and
(\ref{Eq:fit2}) (lower panel) for different choices of
$t_{\rm start}$, spanning from Mar-01-2000 to Dec-01-2000.
The dotted lines show the predicted future trajectories. One see that the
fits are robust with respect to different starting date.}
\label{Fig:all}
\end{figure}

\begin{figure}
\begin{center}
\epsfig{file=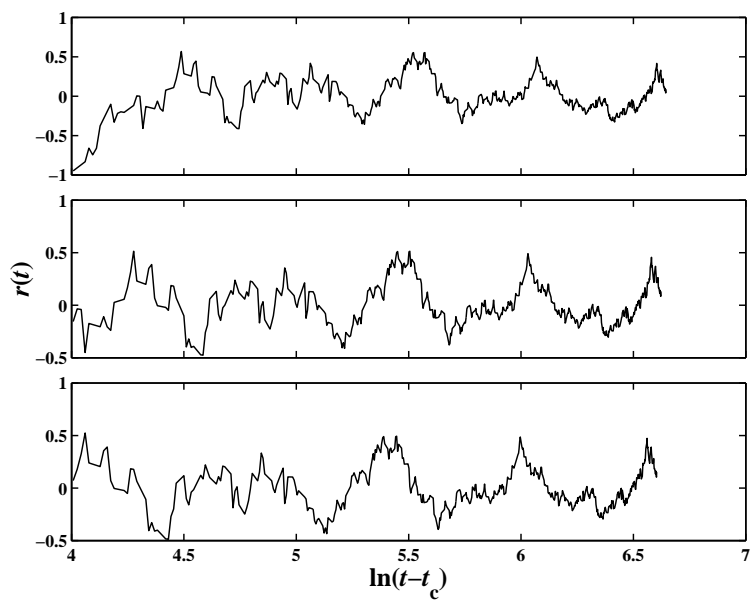,width=10cm, height=8cm}
\end{center}
\caption{The residuals $r(t)$ defined in Eq.~(\ref{Eq:r}) as a
function of $\ln(t-t_c)$. The three plots from top
to bottom correspond respectively to $t_c=$ 15-Jul-2000,
$t_c=$ 01-Aug-2000 and
$t_c=$ 15-Aug-2000.}
\label{Fig:2311}
\end{figure}

\begin{figure}
\begin{center}
\epsfig{file=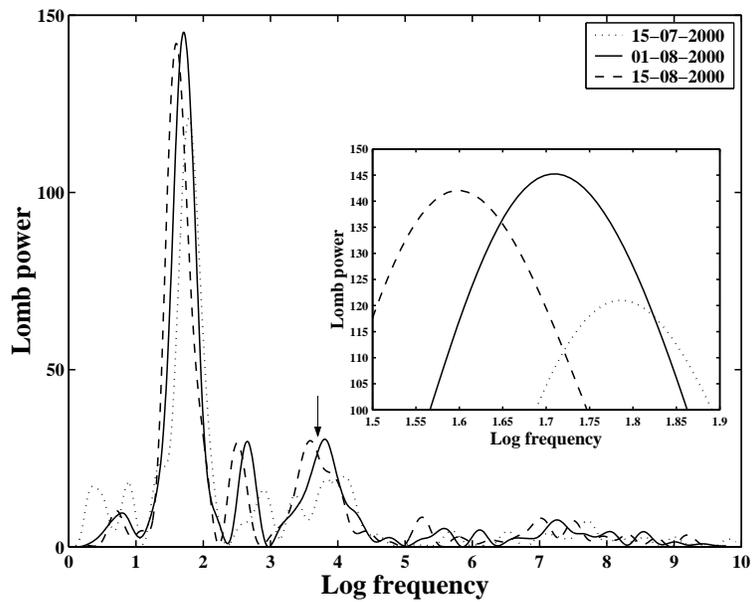,width=10cm, height=8cm}
\end{center}
\caption{Lomb periodograms of the residuals $r(t)$ shown in
Fig.~\ref{Fig:2311}. The highest Lomb peaks are very significant.
The second harmonics is also visible and is indicated by a
downward pointing arrow. The inset shows a magnification of the
Lomb periodogram in the neighborhood of the stronger peaks.}
\label{Fig:2312}
\end{figure}

\begin{figure}
\begin{center}
\epsfig{file=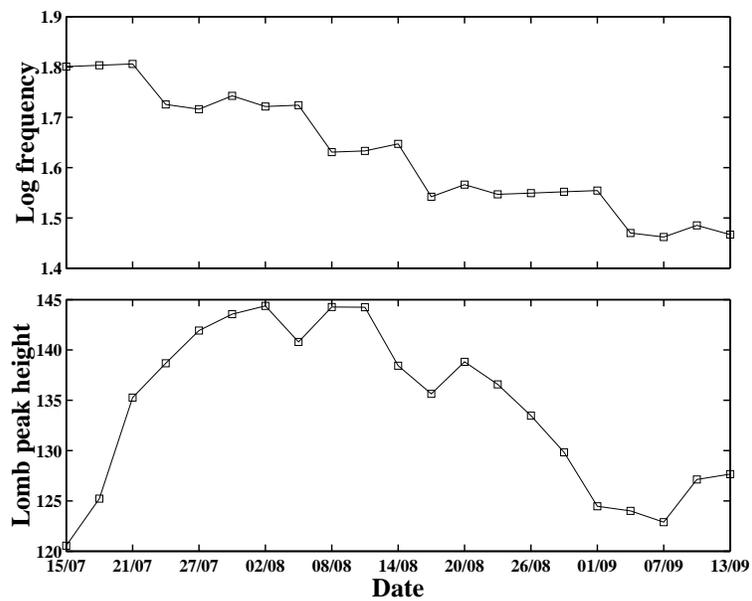,width=10cm, height=8cm}
\end{center}
\caption{Amplitude of the highest Lomb peaks (lower panel) and their associated
log-frequencies (upper panel) obtained by apply the parametric
detrending approach of section \ref{paradet}
for 21 different critical $t_c$ evenly spaced in the time interval
from 15-Jul-2000 to 13-Sep-2000.}
\label{Fig:2313}
\end{figure}

\begin{figure}
\begin{center}
\epsfig{file=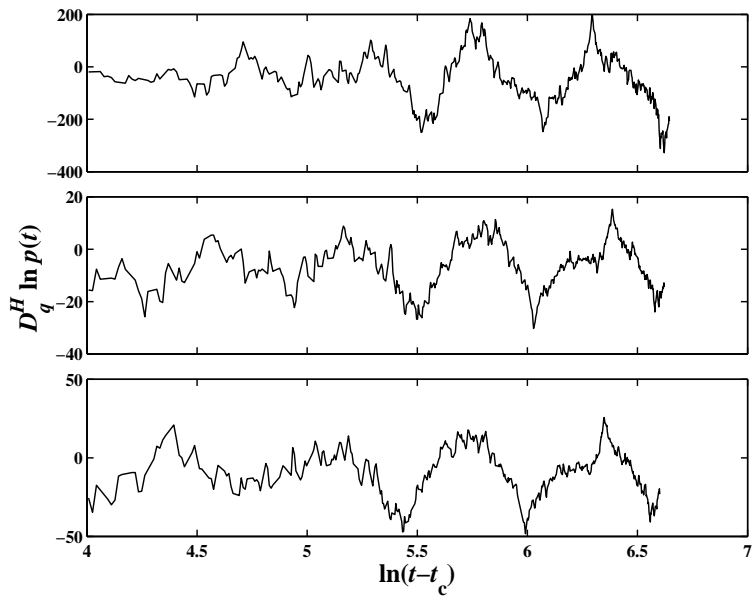,width=10cm, height=8cm}
\end{center}
\caption{The $(H,q)$-derivatives of the logarithm of the S\&P500 index
as a function of $\ln(t-t_c)$ for three value of $t_c$:
15-Jul-2000 with $H = 0$ and $q = 0.8$ in the top panel,
02-Aug-2000 with $H = 0.5$ and $q = 0.7$ in the mid panel, and
16-Aug-2000 with $H = 0.4$ and $q = 0.7$ in the bottom panel.}
\label{Fig:2321}
\end{figure}

\begin{figure}
\begin{center}
\epsfig{file=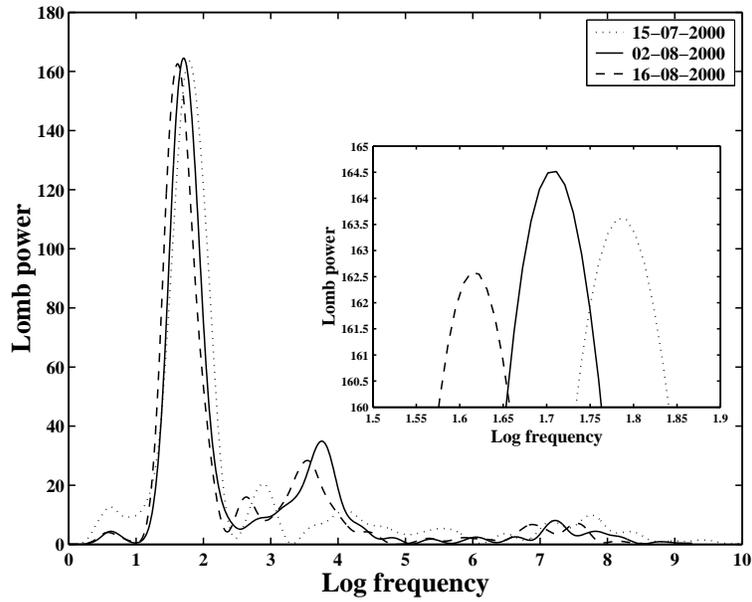,width=10cm, height=8cm}
\end{center}
\caption{Lomb periodograms of the $(H,q)$-derivatives shown in
Fig.~\ref{Fig:2321} for three different value of $t_c$.
The highest Lomb peaks are even more
significant than those in Fig.~\ref{Fig:2312}. The inset shows the magnified
Lomb peaks in the neighborhood of the maxima.}
\label{Fig:2322}
\end{figure}

\begin{figure}
\begin{center}
\epsfig{file=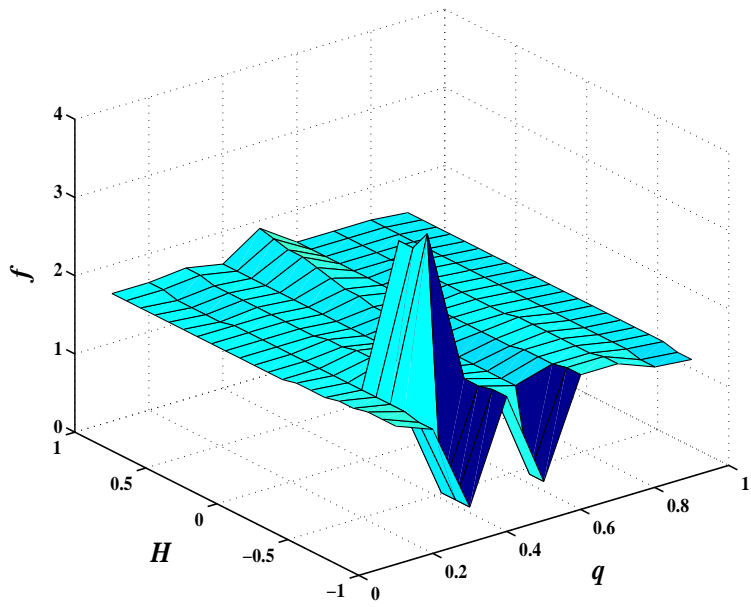,width=10cm, height=8cm}
\end{center}
\caption{Log-frequency $f$ as a function of $H$ and $q$ for
$t_c=$ 16-Aug-2000. The plateau at $f = 1.62 \pm 0.07$ is a signature
of the robustness of the detection of log-periodicity by the
$(H,q)$-analysis. }
\label{Fig:2323}
\end{figure}

\begin{figure}
\begin{center}
\epsfig{file=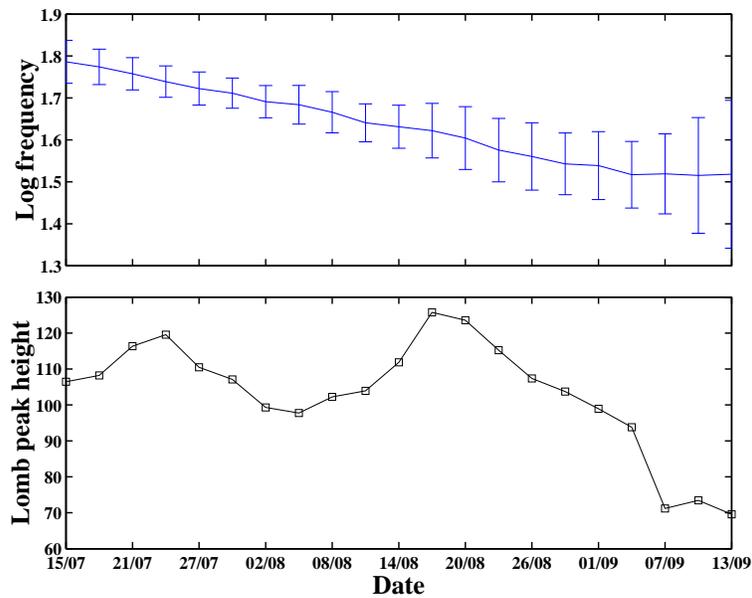,width=10cm, height=8cm}
\end{center}
\caption{The highest Lomb peaks (lower panel) and their associated
log-frequencies (upper panel) obtained by apply the non-parametric
$(H,q)$-analysis for 21 different critical $t_c$ evenly spaced
shown as the abscissa. The Lomb peak
height and the log-frequency
are obtained by averaging all Lomb periodograms over
all pairs $(H, q)$ defined by scanning $H$ from $-1$ to 1
with spacing $0.1$
and $q$ from $0.1$ to $0.9$ with spacing $0.1$.}
\label{Fig:2324}
\end{figure}

\begin{figure}
\begin{center}
\epsfig{file=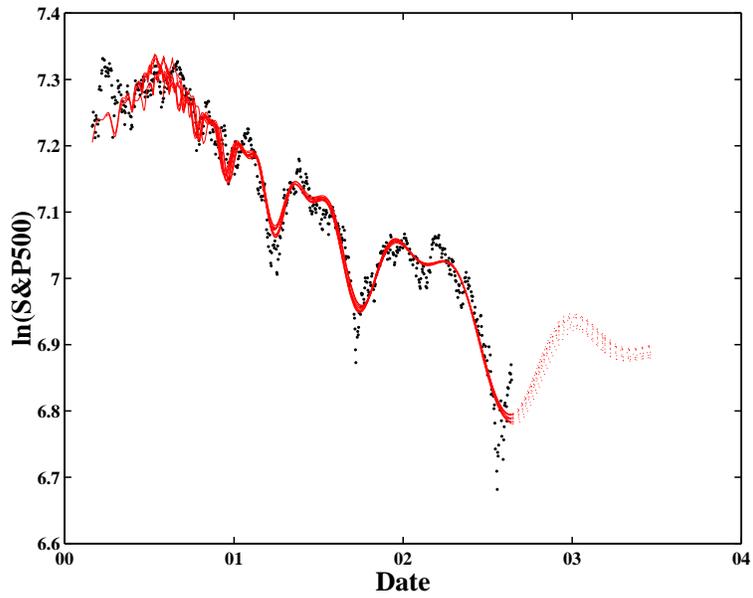,width=10cm, height=8cm}
\end{center}
\caption{All the fitted functions using Eq. (\ref{Eq:fit4}). The
dotted lines show the predicted future trajectories. One sees that
the fits are quite robust with respect to different starting date
$t_{\mathtt{start}}$ from Mar-01-2000 to Dec-01-2000.}
\label{Fig:AllFit4}
\end{figure}

\begin{figure}
\begin{center}
\epsfig{file=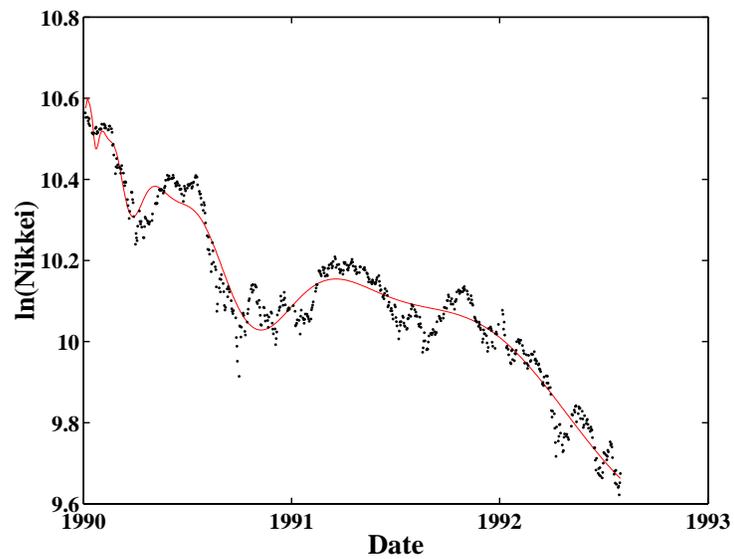,width=10cm, height=8cm}
\end{center}
\caption{Fit of the 2.6 first years of the Nikkei index anti-bubble
from Jan-01-1990 to
Jul-31-1992 by Eq. (\ref{Eq:fit4}) to test for the importance of a
second harmonics. The parameter
values are $t_c=$ 27-Dec-1989, $\alpha=0.42$, $\omega=5.1$,
$\phi_1=5.60$, $\phi_2=1.80$, $A=10.72$, $B=-0.051$, $C=0.0096$
and $D=0.0034$. The r.m.s. error of this fit is $\chi = 0.0457$.}
\label{Fig:Nikkei2}
\end{figure}

\begin{figure}
\begin{center}
\epsfig{file=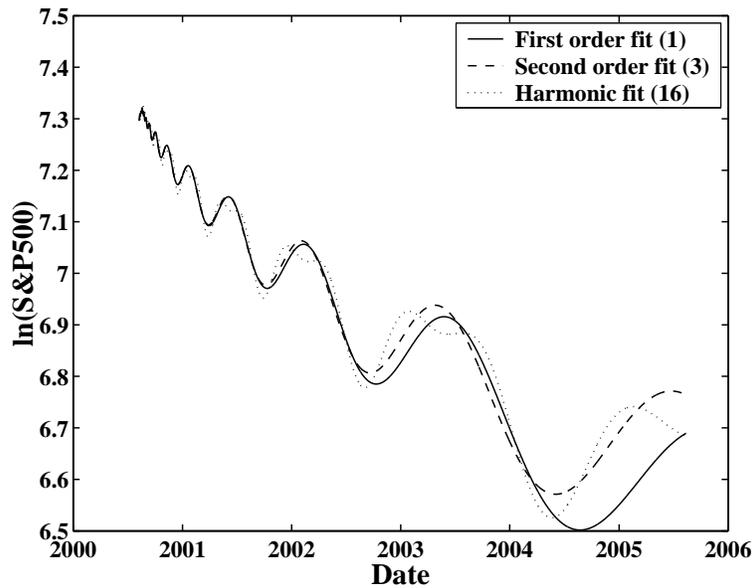,width=10cm, height=8cm}
\end{center}
\caption{Prediction of a change of regime from the first order
formula (\ref{Eq:fit1}) to the
second order formula (\ref{Eq:fit2}). For the first order
formula (\ref{Eq:fit1}), we use
the parameters corresponding to $t_c=$
Aug-01-2000 given in Table \ref{Tb}.
The second order
fit (\ref{Eq:fit2}) uses the same parameter values as the first-order formula
with the addition that $\Delta_t$ and $\Delta_{\omega}$ are fixed to
the values determined the corresponding fit to the Nikkei
index extending over 5 years of data. The crossover between
the two formulas is predicted
to occur in the first semester of 2004. We also show for comparison the fit
using expression (\ref{Eq:fit4}) taking into account the second
harmonics at the angular log-frequency $2\omega$, with
the parameters given in Table \ref{Tb2}.
}
\label{Fig:301}
\end{figure}

The third important point is the improvement in the quality of the fits and
therefore in the potential for predictions when adding the harmonics at
$2 \omega$, as shown
in figure \ref{Fig:AllFit4} compared with figure \ref{Fig:all}. Figure
\ref{Fig:all} suggests a local maximum of the S\&P500 index around the end of the
first quarter of 2003, while figure \ref{Fig:AllFit4} refines this prediction
by seeing an earlier peak before but close to the end of 2002. These two
predictions are not in contradiction: the prediction of figure
\ref{Fig:AllFit4}
shows that the oscillatory structure of S\&P500 index implies several
ups and downs
in the coming year, with a tendency to appreciate for a while before going down
again by the end of 2003.

\begin{figure}
\begin{center}
\epsfig{file=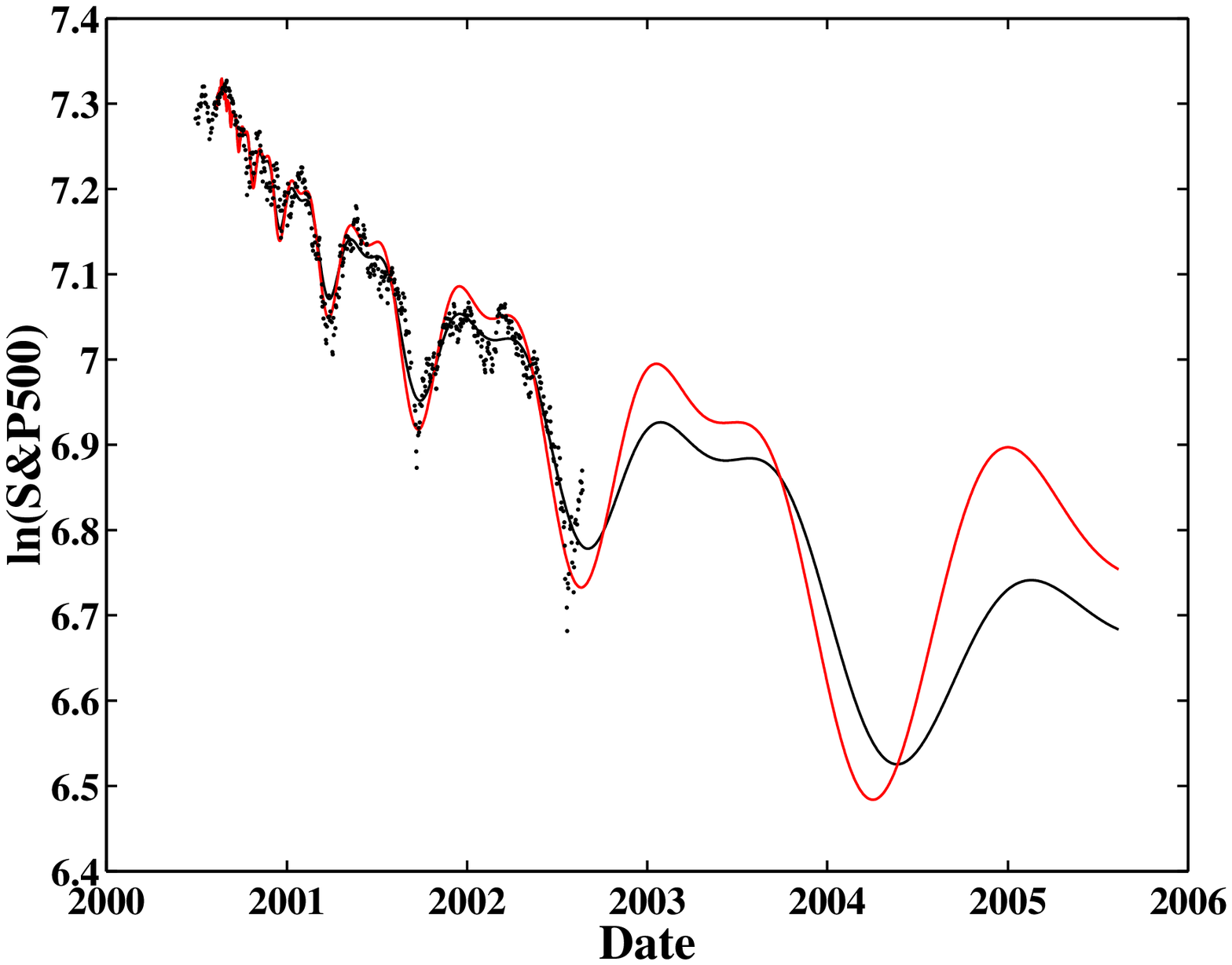,width=10cm, height=8cm}
\end{center}
\caption{Combination of the effect of the second-order formula
(\ref{Eq:fit2}) with the impact of the second harmonics described by
expression (\ref{Eq:fit4}) (see text for explanations).
The corresponding fit (thin continuous line) is compared with expression
(\ref{Eq:fit4}) (second harmonic effect only in thick continuous line)
and extrapolates them up to close to the end of 2006.
}
\label{Fig302}
\end{figure}

Ideally, we would like to combine the effect of the second-order formula
(\ref{Eq:fit2}) with the impact of the second harmonics described by
expression (\ref{Eq:fit4}). We do this by adding to (\ref{Eq:fit4})
a term with the same structure as the one proportional to $C_2$ in (\ref{Eq:fit2})
with $\omega$ replaced by $2 \omega$, again fixing $\Delta_t$
and $\Delta_{\omega}$ at the values determined from the Nikkei index.
Figure \ref{Fig302} compares the result of this fit with that with
(\ref{Eq:fit4}) and their extropolation up to close to the end of 2006.
These two curves provide a sense of the future directions of the S\&P500 index
and their probable degree of variability.

\section{Concluding remarks}
\label{s1:concl}

The growing awareness in 2002 of the crisis in the American
financial system is reminiscent of the starting point of Japan's
massive financial bubble burst more than 10 years before and of
the intertwining of the bad debts and bad performance of banks
whose capital is invested in the shares of other banks, thus
creating the potential for a catastrophic cascade of bankrupts.
Japan has rediscovered before the US the faults of the 19th
century financial system in the US in which stock markets were so
much intertwined with their overall banking financial system, that
busts and bursts occurred more than once every decade, with firms
losing their credit lines and workers and consumers their savings
and often their employment. It is often said that the 1930s
depression was the last of the stock market and bank-induced
economic collapses. The growing fuzziness between financial
banking systems and stock markets, in part due to the innovations
in information technology, has re-created the climate for stronger
bubbles and more pronounced losses of confidence leading to 
long-lived bearish regimes possibly nucleating depressions.

A big problem is that, in the collapse following them,
policy interventions such as lowering interest rates, reducing
taxes, government spending packages and any measure 
to restore investors' confidence may be much less
effective, as discovered with the Japanese so-called liquididy
trap, a process in which government and the central bank policy
becomes essentially useless. In addition, loss of confidence by
investors (for instance following the frauds in accounting in the US)
may lead to a non-negligible
cost to the overall economy \cite{Brookings}, providing a positive
feedback reinforcing the bearish climate.

We have proposed that the trajectories of the US and Japanese
stock markets could be understood in large part by
taking into account imitative and herding mechanisms, both stemming
possibly from rational or irrational behaviors. A key
ingredient entering probably in the imitative and herding processes is
the phenomenon of investor confidence. It has recently been argued \cite{Stout}
that investor confidence can be understood far better 
if one assumes not that investors
have rational expectations, but that they have what economists
call ``adaptive expectations.'' Individuals with rational
expectations predict others' behavior by focusing on their
external incentives and constraints. In contrast, individuals
with adaptive expectations predict others' behavior (including
possibly the behavior of such an abstract ``other'' as the stock
market) by extrapolating from the past. In addition, confidence and trust
in the market have been shown to be subjected to
history effects \cite{Stout}. We believe
that these behavioral traits provide fundamental
roots underlying the validity of our analysis which, ultimately, can be
viewed similarly 
as nothing but a (rather complex nonlinear) extrapolation of the past.

Our theory does not describe the common ``stationary'' evolution
of the stock market, but rather
is specifically tailored for identifying ``monsters'' or anomalies 
(bubbles and their end) and for classifying their agonies. We claim
that these agonies (the anti-bubbles) are mostly shaped by
collective effects between economic and stock market agents, with their
imitation and confidence (and lack thereof) idiosynchracies. Ultimately,
our description must leave place to a recovery of the fundamental pricing principles
(and of possible emergences of new speculative bubbles)
but, before this, it describes the way by which collective effects control
in large part the processes towards this recovery.

The present study complements and makes more precise a previous one
focusing at longer times scales, 
based on three pieces of evidence, namely the growth over long
time scales of population, gross national product and stock market
indices \cite{superbubble,bookcrash}, which issued a prediction that
starting around 1999, a 5 to 10 years consolidation of
international stock markets will occur, allowing a purge after the
over-aggressive appetite of the preceding decade. For more than
the last two years, this prediction has been born out. The present
study confirms this impression that the US stock market is not yet 
on the verge of recovery.

With its
extraordinary and unparalleled growth, its ensuing decade-long
absence of growth, its crowded land, its aging population, is
Japan a precursor of the new regime that mankind has to shift to,
as discussed in \cite{superbubble,bookcrash}? It seems that the
present qualification of the US market to be in an anti-bubble
phase is entirely in line with these predictions. From a larger
perspective and at the horizon of the end of the first half of this century, the
behavior of these stock markets raise the following question:
shall we learn the lessons of previous bubbles and
crashes/depressions and shall we be able to transit to a
qualitatively different organization of economic and cultural
exchanges before the fundamental limitations of a finite earth and
limited human intelligence set in?

\bigskip
{\bf Acknowledgments:} We acknowledge stimulating discussions with D. Darcet
and thank him for pointing out the similarity between the Nikkei and
the S\&P500 with a time-shift of about 11 years. We thank D. Stauffer
for critical comments on the manuscript.
This work was partially supported by
NSF-DMR99-71475 and the James S. Mc Donnell Foundation 21st
century scientist award/studying complex system.

\pagebreak

\clearpage

\begin{table}
\begin{center}
\caption{\label{Tb} Fitted parameters using equations
(\ref{Eq:fit1}) and (\ref{Eq:fit2}) on the S\&P500 index. The
parameters with subscript 0 correspond to fits using equation
(\ref{Eq:fit1}) together with (\ref{Eq:tau1}). The subscripts 1
and 2 refer to fits with expressions (\ref{Eq:fit1}) and
(\ref{Eq:fit2}) respectively together with (\ref{Eq:tau3}). The
rows of $\chi_0$, $\chi_1$ and $\chi_2$ present the standard
deviations (r.m.s.) of the residuals for different fits. Note that
the fits with the first-order formula (\ref{Eq:fit1}) with
definition (\ref{Eq:tau1}) are unstable and sensitive to
$t_{\rm{start}}$. Comparing parameters with subscripts 1 and 2,
one can see that applying (\ref{Eq:tau3}) improves significantly
the quality of the fits. With this new symmetric definition
(\ref{Eq:tau3}), we find that the first-order fits using
(\ref{Eq:fit1}) are quite robust and close to the fits with the
second-order formula (\ref{Eq:fit2}). The {\textbf{bold}} values
are discussed in the text.}
\medskip
\begin{tabular}{ccccccccccc}
\hline\hline
$t_{\rm{start}}$&01/03&01/04&{\bf01/05}&{\bf01/06}&{\bf01/07}&{\bf01/0
8}&{\bf01/09}&01/10&01/11&01/12\\\hline
$t_{c,0}$&02/28&01/24&04/05&03/16&06/21&07/19&08/15&07/16&08/13&09/12\\
$t_{c,1}$&07/12&07/18&{\bf08/06}&{\bf08/06}&{\bf08/06}&{\bf08/18}&{\bf
08/10}&08/10&08/17&09/18\\
$t_{c,2}$&07/13&07/19&08/05&08/06&08/13&08/18&08/11&08/10&08/15&09/15\\
$\chi_0\times{100}$&3.969&3.790&3.752&3.605&3.423&3.303&3.277&3.363&3.
369&3.406\\
$\chi_1\times{100}$&3.431&3.317&{\bf3.187}&{\bf3.217}&{\bf3.229}&{\bf3
..218}&{\bf3.277}&3.307&3.370&3.407\\
$\chi_2\times{100}$&3.415&3.315&3.182&3.215&3.209&3.218&3.276&3.307&3.
370&3.407\\
$\alpha_0$&1.00&1.00&1.00&0.83&0.71&0.67&0.67&0.70&0.75&0.76\\
$\alpha_1$&0.80&0.74&{\bf0.71}&{\bf0.70}&{\bf0.69}&{\bf0.68}&{\bf0.66}
&0.73&0.75&0.77\\
$\alpha_2$&0.76&0.76&0.70&0.70&0.68&0.68&0.66&0.73&0.75&0.76\\
$\omega_0$&14.23&15.32&13.12&13.77&12.09&11.00&10.12&11.20&10.19& 9.37\\
$\omega_1$&10.79&11.01&{\bf10.30}&{\bf10.35}&{\bf10.50}&{\bf10.04}&{\b
f10.31}&10.30&10.06& 9.17\\
$\omega_2$&11.33&10.82&10.37&10.38&10.22&10.03&10.27&10.29&10.11& 9.29\\
$\phi_0$&3.49&4.99&2.03&0.59&4.16&5.34&5.17&4.01&1.55&4.09\\
$\phi_1$&0.15&2.10&3.78&3.46&5.74&5.72&3.86&3.91&5.53&2.29\\
$\phi_2$&6.28&3.31&3.34&3.27&1.35&5.80&4.12&3.97&2.03&4.67\\
$A_1$&7.33&7.33&7.33&7.33&7.33&7.32&7.34&7.31&7.30&7.27\\
$A_2$&7.33&7.33&7.33&7.33&7.33&7.32&7.33&7.31&7.30&7.28\\
$B_1\times{1000}$&-2.10&-3.10&-4.10&-4.20&-4.38&-4.92&-5.66&-3.40&-2.78&-2.41\\
$B_2\times{1000}$&-2.85&-2.87&-4.23&-4.27&-4.88&-4.89&-5.50&-3.37&-2.78&-2.54\\
$C_1\times{1000}$&0.47&-0.68&0.92&0.94&-0.99&1.13&1.25&0.82&0.70&-0.67\\
$C_2\times{1000}$&-0.62&-0.64&0.94&0.95&-1.10&1.13&1.22&0.81&-0.70&0.70\\
$\Delta{t}$&12128&24032&35135&93922&64998&94620&100000&97552&94581&98337\\
$\Delta{\omega}$&0.00&4.92&6.04&8.39&0.11&2.79&0.00&0.00&0.26&9.98\\\hline\hline
\end{tabular}
\end{center}
\end{table}

\clearpage

\begin{table}
\begin{center}
\caption{\label{Tb2} Parameters of the fit with equation
(\ref{Eq:fit4}) with (\ref{Eq:tau3}) of the S\&P500 index for
different $t_{\rm start}$. The fit residuals are strongly
reduced compared with the fits shown in Table \ref{Tb}. The
{\textbf{bold}} columns correspond to the values of $t_{\rm start}$
giving basically the same values for $t_c, \alpha$ and $\omega$.}
\medskip
\begin{tabular}{ccccccccccc}
\hline\hline
$t_{\rm{start}}$&01/03&01/04&01/05&{\bf01/06}&{\bf01/07}&{\bf01/08}&{\
bf01/09}&01/10&01/11&01/12\\\hline
$t_{c}$&06/15&07/15&07/16&08/04&08/12&08/21&08/04&06/14&08/10&06/04\\
$\chi\times{100}$&3.009&2.858&2.713&2.711&2.680&2.690&2.661&2.700&2.712&2.687\\
$\alpha$&0.79&0.68&0.65&0.65&0.61&0.61&0.54&0.60&0.57&0.57\\
$\omega$&12.12&11.56&11.77&10.70&10.74&10.26&10.97&12.24&10.86&12.49\\
$\phi_1$&3.88&5.01&3.76&1.41&1.39&4.52&2.89&6.28&0.59&1.35\\
$\phi_2$&4.90&0.66&4.39&2.98&2.80&2.84&5.83&0.24&1.20&2.88\\
$A$&7.33&7.34&7.34&7.34&7.34&7.33&7.38&7.39&7.35&7.40\\
$B\times{1000}$&-2.19&-4.68&-5.59&-5.77&-7.60&-7.36&-12.42&-8.38&-10.3
4&-10.67\\
$C\times{1000}$&0.44&-0.92&-1.09&1.21&1.54&1.58&-2.29&-1.44&2.03&1.79\\
$D\times{1000}$&0.22&0.47&0.59&-0.53&-0.77&-0.71&-1.14&-0.77&-1.03&-0.96\\
\hline\hline
\end{tabular}
\end{center}
\end{table}

\clearpage

\begin{table}
\begin{center}
\caption{\label{Tb3} Likelihood-ratio test of the hypothesis that
$D\neq 0$ in Eq.~(\ref{Eq:fit4}). The beginning of the data set
for fit is set to be fixed at 09-Aug-2000. The end of the data set
varies from 01-Jan-2001 to 24-Aug-2002. The $n$ column gives the
number of the data set for each fit. The confidence level
quantified by ${\rm{Proba}}$ decreases on average with $n$. There
is no doubt that the $H_1$ hypothesis of $D\neq 0$ can not be
rejected for all cases in the $99.8\%$ confidence level.}
\medskip
\begin{tabular}{ccccccccccc}
\hline\hline
$t_{\rm{last}}$&$n$&$\sigma_1$&$\sigma_0$&$T$&${\rm{Proba}}$\\\hline
01/01/01&101&0.0235&0.0283&18.7&$0.0015\%$\\
01/31/01&121&0.0310&0.0381&24.8&$<10^{-4}\%$\\
03/02/01&142&0.0341&0.0440&36.2&$<10^{-4}\%$\\
04/01/01&163&0.0578&0.0616&10.3&$0.13\%$\\
05/01/01&183&0.0831&0.0934&21.3&$0.0004\%$\\
05/31/01&204&0.0949&0.1026&15.9&$0.0067\%$\\
06/30/01&225&0.0984&0.1064&17.7&$0.0026\%$\\
07/30/01&245&0.1022&0.1260&51.2&$<10^{-4}\%$\\
08/29/01&267&0.1087&0.1544&93.9&$<10^{-4}\%$\\
09/28/01&284&0.1332&0.1897&100.4&$<10^{-4}\%$\\
10/28/01&305&0.1817&0.2181&55.7&$<10^{-4}\%$\\
11/27/01&325&0.1855&0.2512&98.4&$<10^{-4}\%$\\
12/27/01&346&0.1971&0.2552&89.4&$<10^{-4}\%$\\
01/26/02&365&0.1987&0.2659&106.2&$<10^{-4}\%$\\
02/25/02&385&0.2055&0.2895&132.0&$<10^{-4}\%$\\
03/27/02&407&0.2329&0.3274&138.6&$<10^{-4}\%$\\
04/26/02&428&0.2363&0.3353&149.7&$<10^{-4}\%$\\
05/26/02&448&0.2482&0.3456&148.3&$<10^{-4}\%$\\
06/25/02&469&0.2573&0.3515&146.4&$<10^{-4}\%$\\
07/25/02&489&0.3170&0.4867&209.6&$<10^{-4}\%$\\
08/24/02&510&0.3582&0.5339&203.6&$<10^{-4}\%$\\
\hline\hline
\end{tabular}
\end{center}
\end{table}

\clearpage

\clearpage

\clearpage

\clearpage

\clearpage

\clearpage

\clearpage

\clearpage

\clearpage

\clearpage

\clearpage

\clearpage

\clearpage


\begin{thebibliography}{}

\bibitem{Bender} Bender C, Orszag S.A.,
  {\it Advanced Mathematical Methods for Scientists and Engineers.}
McGraw-Hill, New York  (1978), page 147.

\bibitem{Erzan} A. Erzan, Finite q-differences and the discrete
renormalization group, Phys. Lett. A 225, 235-238 (1997).

\bibitem{ErzEck} Erzan, A. and J.P. Eckmann, q-analysis of Fractal
Sets, Phys. Rev. Lett. 87, 3245-3248 (1997).

\bibitem{Brookings} C. Graham, R. Litan and S. Sukhtanker,
Cooking the Books: The Cost to the Economy, Brookings Policy Brief 106, 02/08 (2002)
($http://www.brookings.edu/comm/policybriefs/pb106.htm$).

\bibitem{Holden} Holden K, Peel DA and Thompson JL, Economic
Forecasting: An Introduction (Cambridge University Press,
Cambridge, 1990) pp.59.

\bibitem{fmp} Huang, Y., A. Johansen, M.W. Lee, H. Saleur and
D. Sornette, Artifactual log-periodicity in finite size data:
Relevance for earthquake aftershocks, J. Geophys. Res. 105,
25451-25471 (2000).

\bibitem{Idesor} K. Ide and D. Sornette,
Oscillatory Finite-Time Singularities in Finance, Population and Rupture, 
Physica A  307 (1-2), 63-106 (2002)

\bibitem{JohSorLed99} A. Johansen, D. Sornette and O. Ledoit, Predicting
financial crashes using discrete scale invariance, Journal of Risk
1, 5-32 (1999).

\bibitem{CriCrash00} A. Johansen, O. Ledoit and D. Sornette,
Crashes as critical points, International Journal of Theoretical
and Applied Finance Vol. 3, No. 2 219-255 (2000).

\bibitem{CriCrash99} A. Johansen and D. Sornette, Critical Crashes,
Risk, Vol 12, No. 1, p.91-94 (1999).

\bibitem{Nikkei99} A. Johansen and D. Sornette, Financial
``anti-bubbles'': Log-periodicity in Gold and Nikkei collapses,
Int. J. Mod. Phys. C 10(4), 563-575 (1999).

\bibitem{JS00} A. Johansen and D. Sornette,
Evaluation of the quantitative prediction of a trend reversal on
the Japanese stock market in 1999, Int. J. Mod. Phys. C Vol. 11 (2),
359-364 (2000).

\bibitem{nasdaqlo} A. Johansen and D. Sornette,
The Nasdaq crash of April 2000: Yet another example of log-periodicity
  in a speculative bubble ending in a crash,
  European Physical Journal B 17, 319-328 (2000).

\bibitem{superbubble} A. Johansen and D. Sornette,
Finite-time singularity in the dynamics of the world population and
economic indices,
Physica A 294 (3-4), 465-502 (2001).

\bibitem{JSH} A. Johansen, D. Sornette and A.E. Hansen,
Punctuated vortex coalescence and
discrete scale invariance in two-dimensional turbulence, Physica D 138, 302-315 (2000).

\bibitem{JSL99} A. Johansen, D. Sornette and O. Ledoit,
Predicting Financial Crashes using discrete scale invariance,
Journal of Risk 1 (4), 5-32 (1999)

\bibitem{Press} Press, W., S. Teukolsky, W. Vetterling and B.
Flannery, Numerical Recipes in FORTRAN: The Art of Scientific
Computing (Cambridge University, Cambridge, 1996).

\bibitem{Rao} Rao C, Linear statistical Inference and Its
Applications (New York, Wiley, 1965) ch 6, section 6e.3.

\bibitem{SorDSI} D. Sornette, Discrete scale invariance and complex
dimensions, Physics Reports 297, 239-270 (1998).

\bibitem{bookcrash} D. Sornette, Why Stock Markets Crash
(Critical Events in Complex Financial Systems),
Princeton University Press, Princeton, NJ, 2002, in press.

\bibitem{SJ97} D. Sornette and A. Johansen, Large financial
crashes, Physica A 245, N3-4, 411-422 (1997).

\bibitem{SorJoh01} D. Sornette and A. Johansen, Significance of
log-periodic precursors to financial crashes, Quantitative Finance
1, 452-471 (2001).

\bibitem{SJB96} D. Sornette, A. Johansen and J.-P. Bouchaud,
Stock market crashes, Precursors and Replicas, J.Phys.I France 6,
167-175  (1996)

\bibitem{Stout} L.A. Stout, The Investor Confidence Game, 
working paper available
$http://papers.ssrn.com/paper.taf?abstract\_id=322301$

\bibitem{Wilson} Wilson, K.G.,
Problems in Physics with many scales of length,
Scientific American 241 (2), 158-179 (1979).

\bibitem{Zhou02a} W.-X. Zhou and D. Sornette, Statistical
significance of periodicity and log-periodicity with heavy-tailed
correlated noise, Int. J. Mod. Phys. C 13 (2), 137-170 (2002).

\bibitem{turb1} W.-X. Zhou and D. Sornette,
Evidence of Intermittent Cascades from Discrete Hierarchical
Dissipation in Turbulence, Physica D 165, 94-125 (2002).

\bibitem{HqARup} W.-X. Zhou and D. Sornette, Generalized q-Analysis of
Log-Periodicity: Applications to Critical Ruptures, Phys. Rev. E,
in press, http://arXiv.org/abs/cond-mat/0201458.

\bibitem{HqACrash} W.-X. Zhou and D. Sornette, Non-Parametric Analyses of
Log-Periodic Precursors to Financial Crashes, preprint,
http://arXiv.org/abs/cond-mat/0205531.

\bibitem{turb2} W.-X. Zhou, D. Sornette and V. Pisarenko,
New Evidence of Discrete Scale Invariance in the Energy Dissipation of
Three-Dimensional Turbulence: Correlation Approach and Direct
Spectral Detection, submitted to Physical Review E  (2002).
(http://arXiv.org/abs/cond-mat/0208347)

\bibitem{BubbleTests} W.-X. Zhou and D. Sornette,
Positive Hazard Rate and Log-Periodicity, in preparation.


\end{thebibliography}
\end{document}